\documentstyle[12pt]{article}
\begin{document}
\begin{flushright}
Preprint SSU-HEP-98/03\\
Samara State University
\end{flushright}

\vspace{30mm}

\immediate\write16{<<WARNING: LINEDRAW macros work with emTeX-dvivers
		    and other drivers supporting emTeX \special's
		    (dviscr, dvihplj, dvidot, dvips, dviwin, etc.) >>}
\newdimen\Lengthunit	   \Lengthunit	= 1.5cm
\newcount\Nhalfperiods	   \Nhalfperiods= 9
\newcount\magnitude	   \magnitude = 1000

\catcode`\*=11
\newdimen\L*   \newdimen\d*   \newdimen\d**
\newdimen\dm*  \newdimen\dd*  \newdimen\dt*
\newdimen\a*   \newdimen\b*   \newdimen\c*
\newdimen\a**  \newdimen\b**
\newdimen\xL*  \newdimen\yL*
\newdimen\rx*  \newdimen\ry*
\newdimen\tmp* \newdimen\linwid*

\newcount\k*   \newcount\l*   \newcount\m*
\newcount\k**  \newcount\l**  \newcount\m**
\newcount\n*   \newcount\dn*  \newcount\r*
\newcount\N*   \newcount\*one \newcount\*two  \*one=1 \*two=2
\newcount\*ths \*ths=1000
\newcount\angle*  \newcount\q*	\newcount\q**
\newcount\angle** \angle**=0
\newcount\sc*	  \sc*=0

\newtoks\cos*  \cos*={1}
\newtoks\sin*  \sin*={0}

\catcode`\[=13

\def\rotate(#1){\advance\angle**#1\angle*=\angle**
\q**=\angle*\ifnum\q**<0\q**=-\q**\fi
\ifnum\q**>360\q*=\angle*\divide\q*360\multiply\q*360\advance\angle*-\q*\fi
\ifnum\angle*<0\advance\angle*360\fi\q**=\angle*\divide\q**90\q**=\q**
\def\sgcos*{+}\def\sgsin*{+}\relax
\ifcase\q**\or
 \def\sgcos*{-}\def\sgsin*{+}\or
 \def\sgcos*{-}\def\sgsin*{-}\or
 \def\sgcos*{+}\def\sgsin*{-}\else\fi
\q*=\q**
\multiply\q*90\advance\angle*-\q*
\ifnum\angle*>45\sc*=1\angle*=-\angle*\advance\angle*90\else\sc*=0\fi
\def[##1,##2]{\ifnum\sc*=0\relax
\edef\cs*{\sgcos*.##1}\edef\sn*{\sgsin*.##2}\ifcase\q**\or
 \edef\cs*{\sgcos*.##2}\edef\sn*{\sgsin*.##1}\or
 \edef\cs*{\sgcos*.##1}\edef\sn*{\sgsin*.##2}\or
 \edef\cs*{\sgcos*.##2}\edef\sn*{\sgsin*.##1}\else\fi\else
\edef\cs*{\sgcos*.##2}\edef\sn*{\sgsin*.##1}\ifcase\q**\or
 \edef\cs*{\sgcos*.##1}\edef\sn*{\sgsin*.##2}\or
 \edef\cs*{\sgcos*.##2}\edef\sn*{\sgsin*.##1}\or
 \edef\cs*{\sgcos*.##1}\edef\sn*{\sgsin*.##2}\else\fi\fi
\cos*={\cs*}\sin*={\sn*}\global\edef\gcos*{\cs*}\global\edef\gsin*{\sn*}}\relax
\ifcase\angle*[9999,0]\or
[999,017]\or[999,034]\or[998,052]\or[997,069]\or[996,087]\or
[994,104]\or[992,121]\or[990,139]\or[987,156]\or[984,173]\or
[981,190]\or[978,207]\or[974,224]\or[970,241]\or[965,258]\or
[961,275]\or[956,292]\or[951,309]\or[945,325]\or[939,342]\or
[933,358]\or[927,374]\or[920,390]\or[913,406]\or[906,422]\or
[898,438]\or[891,453]\or[882,469]\or[874,484]\or[866,499]\or
[857,515]\or[848,529]\or[838,544]\or[829,559]\or[819,573]\or
[809,587]\or[798,601]\or[788,615]\or[777,629]\or[766,642]\or
[754,656]\or[743,669]\or[731,681]\or[719,694]\or[707,707]\or
\else[9999,0]\fi}

\catcode`\[=12

\def\GRAPH(hsize=#1)#2{\hbox to #1\Lengthunit{#2\hss}}

\def\Linewidth#1{\global\linwid*=#1\relax
\global\divide\linwid*10\global\multiply\linwid*\mag
\global\divide\linwid*100\special{em:linewidth \the\linwid*}}

\Linewidth{.4pt}
\def\sm*{\special{em:moveto}}
\def\sl*{\special{em:lineto}}
\let\moveto=\sm*
\let\lineto=\sl*
\newbox\spm*   \newbox\spl*
\setbox\spm*\hbox{\sm*}
\setbox\spl*\hbox{\sl*}

\def\mov#1(#2,#3)#4{\rlap{\L*=#1\Lengthunit
\xL*=#2\L* \yL*=#3\L*
\xL*=\xscale\xL* \yL*=\yscale\yL*
\rx* \the\cos*\xL* \tmp* \the\sin*\yL* \advance\rx*-\tmp*
\ry* \the\cos*\yL* \tmp* \the\sin*\xL* \advance\ry*\tmp*
\kern\rx*\raise\ry*\hbox{#4}}}

\def\rmov*(#1,#2)#3{\rlap{\xL*=#1\yL*=#2\relax
\rx* \the\cos*\xL* \tmp* \the\sin*\yL* \advance\rx*-\tmp*
\ry* \the\cos*\yL* \tmp* \the\sin*\xL* \advance\ry*\tmp*
\kern\rx*\raise\ry*\hbox{#3}}}

\def\lin#1(#2,#3){\rlap{\sm*\mov#1(#2,#3){\sl*}}}

\def\arr*(#1,#2,#3){\rmov*(#1\dd*,#1\dt*){\sm*
\rmov*(#2\dd*,#2\dt*){\rmov*(#3\dt*,-#3\dd*){\sl*}}\sm*
\rmov*(#2\dd*,#2\dt*){\rmov*(-#3\dt*,#3\dd*){\sl*}}}}

\def\arrow#1(#2,#3){\rlap{\lin#1(#2,#3)\mov#1(#2,#3){\relax
\d**=-.012\Lengthunit\dd*=#2\d**\dt*=#3\d**
\arr*(1,10,4)\arr*(3,8,4)\arr*(4.8,4.2,3)}}}

\def\arrlin#1(#2,#3){\rlap{\L*=#1\Lengthunit\L*=.5\L*
\lin#1(#2,#3)\rmov*(#2\L*,#3\L*){\arrow.1(#2,#3)}}}

\def\dasharrow#1(#2,#3){\rlap{{\Lengthunit=0.9\Lengthunit
\dashlin#1(#2,#3)\mov#1(#2,#3){\sm*}}\mov#1(#2,#3){\sl*
\d**=-.012\Lengthunit\dd*=#2\d**\dt*=#3\d**
\arr*(1,10,4)\arr*(3,8,4)\arr*(4.8,4.2,3)}}}

\def\clap#1{\hbox to 0pt{\hss #1\hss}}

\def\ind(#1,#2)#3{\rlap{\L*=.1\Lengthunit
\xL*=#1\L* \yL*=#2\L*
\rx* \the\cos*\xL* \tmp* \the\sin*\yL* \advance\rx*-\tmp*
\ry* \the\cos*\yL* \tmp* \the\sin*\xL* \advance\ry*\tmp*
\kern\rx*\raise\ry*\hbox{\lower2pt\clap{$#3$}}}}

\def\sh*(#1,#2)#3{\rlap{\dm*=\the\n*\d**
\xL*=\xscale\dm* \yL*=\yscale\dm* \xL*=#1\xL* \yL*=#2\yL*
\rx* \the\cos*\xL* \tmp* \the\sin*\yL* \advance\rx*-\tmp*
\ry* \the\cos*\yL* \tmp* \the\sin*\xL* \advance\ry*\tmp*
\kern\rx*\raise\ry*\hbox{#3}}}

\def\calcnum*#1(#2,#3){\a*=1000sp\b*=1000sp\a*=#2\a*\b*=#3\b*
\ifdim\a*<0pt\a*-\a*\fi\ifdim\b*<0pt\b*-\b*\fi
\ifdim\a*>\b*\c*=.96\a*\advance\c*.4\b*
\else\c*=.96\b*\advance\c*.4\a*\fi
\k*\a*\multiply\k*\k*\l*\b*\multiply\l*\l*
\m*\k*\advance\m*\l*\n*\c*\r*\n*\multiply\n*\n*
\dn*\m*\advance\dn*-\n*\divide\dn*2\divide\dn*\r*
\advance\r*\dn*
\c*=\the\Nhalfperiods5sp\c*=#1\c*\ifdim\c*<0pt\c*-\c*\fi
\multiply\c*\r*\N*\c*\divide\N*10000}

\def\dashlin#1(#2,#3){\rlap{\calcnum*#1(#2,#3)\relax
\d**=#1\Lengthunit\ifdim\d**<0pt\d**-\d**\fi
\divide\N*2\multiply\N*2\advance\N*\*one
\divide\d**\N*\sm*\n*\*one\sh*(#2,#3){\sl*}\loop
\advance\n*\*one\sh*(#2,#3){\sm*}\advance\n*\*one
\sh*(#2,#3){\sl*}\ifnum\n*<\N*\repeat}}

\def\dashdotlin#1(#2,#3){\rlap{\calcnum*#1(#2,#3)\relax
\d**=#1\Lengthunit\ifdim\d**<0pt\d**-\d**\fi
\divide\N*2\multiply\N*2\advance\N*1\multiply\N*2\relax
\divide\d**\N*\sm*\n*\*two\sh*(#2,#3){\sl*}\loop
\advance\n*\*one\sh*(#2,#3){\kern-1.48pt\lower.5pt\hbox{\rm.}}\relax
\advance\n*\*one\sh*(#2,#3){\sm*}\advance\n*\*two
\sh*(#2,#3){\sl*}\ifnum\n*<\N*\repeat}}

\def\shl*(#1,#2)#3{\kern#1#3\lower#2#3\hbox{\unhcopy\spl*}}

\def\trianglin#1(#2,#3){\rlap{\toks0={#2}\toks1={#3}\calcnum*#1(#2,#3)\relax
\dd*=.57\Lengthunit\dd*=#1\dd*\divide\dd*\N*
\divide\dd*\*ths \multiply\dd*\magnitude
\d**=#1\Lengthunit\ifdim\d**<0pt\d**-\d**\fi
\multiply\N*2\divide\d**\N*\sm*\n*\*one\loop
\shl**{\dd*}\dd*-\dd*\advance\n*2\relax
\ifnum\n*<\N*\repeat\n*\N*\shl**{0pt}}}

\def\wavelin#1(#2,#3){\rlap{\toks0={#2}\toks1={#3}\calcnum*#1(#2,#3)\relax
\dd*=.23\Lengthunit\dd*=#1\dd*\divide\dd*\N*
\divide\dd*\*ths \multiply\dd*\magnitude
\d**=#1\Lengthunit\ifdim\d**<0pt\d**-\d**\fi
\multiply\N*4\divide\d**\N*\sm*\n*\*one\loop
\shl**{\dd*}\dt*=1.3\dd*\advance\n*\*one
\shl**{\dt*}\advance\n*\*one
\shl**{\dd*}\advance\n*\*two
\dd*-\dd*\ifnum\n*<\N*\repeat\n*\N*\shl**{0pt}}}

\def\w*lin(#1,#2){\rlap{\toks0={#1}\toks1={#2}\d**=\Lengthunit\dd*=-.12\d**
\divide\dd*\*ths \multiply\dd*\magnitude
\N*8\divide\d**\N*\sm*\n*\*one\loop
\shl**{\dd*}\dt*=1.3\dd*\advance\n*\*one
\shl**{\dt*}\advance\n*\*one
\shl**{\dd*}\advance\n*\*one
\shl**{0pt}\dd*-\dd*\advance\n*1\ifnum\n*<\N*\repeat}}

\def\l*arc(#1,#2)[#3][#4]{\rlap{\toks0={#1}\toks1={#2}\d**=\Lengthunit
\dd*=#3.037\d**\dd*=#4\dd*\dt*=#3.049\d**\dt*=#4\dt*\ifdim\d**>10mm\relax
\d**=.25\d**\n*\*one\shl**{-\dd*}\n*\*two\shl**{-\dt*}\n*3\relax
\shl**{-\dd*}\n*4\relax\shl**{0pt}\else
\ifdim\d**>5mm\d**=.5\d**\n*\*one\shl**{-\dt*}\n*\*two
\shl**{0pt}\else\n*\*one\shl**{0pt}\fi\fi}}

\def\d*arc(#1,#2)[#3][#4]{\rlap{\toks0={#1}\toks1={#2}\d**=\Lengthunit
\dd*=#3.037\d**\dd*=#4\dd*\d**=.25\d**\sm*\n*\*one\shl**{-\dd*}\relax
\n*3\relax\sh*(#1,#2){\xL*=\xscale\dd*\yL*=\yscale\dd*
\kern#2\xL*\lower#1\yL*\hbox{\sm*}}\n*4\relax\shl**{0pt}}}

\def\shl**#1{\c*=\the\n*\d**\d*=#1\relax
\a*=\the\toks0\c*\b*=\the\toks1\d*\advance\a*-\b*
\b*=\the\toks1\c*\d*=\the\toks0\d*\advance\b*\d*
\a*=\xscale\a*\b*=\yscale\b*
\rx* \the\cos*\a* \tmp* \the\sin*\b* \advance\rx*-\tmp*
\ry* \the\cos*\b* \tmp* \the\sin*\a* \advance\ry*\tmp*
\raise\ry*\rlap{\kern\rx*\unhcopy\spl*}}

\def\wlin*#1(#2,#3)[#4]{\rlap{\toks0={#2}\toks1={#3}\relax
\c*=#1\l*\c*\c*=.01\Lengthunit\m*\c*\divide\l*\m*
\c*=\the\Nhalfperiods5sp\multiply\c*\l*\N*\c*\divide\N*\*ths
\divide\N*2\multiply\N*2\advance\N*\*one
\dd*=.002\Lengthunit\dd*=#4\dd*\multiply\dd*\l*\divide\dd*\N*
\divide\dd*\*ths \multiply\dd*\magnitude
\d**=#1\multiply\N*4\divide\d**\N*\sm*\n*\*one\loop
\shl**{\dd*}\dt*=1.3\dd*\advance\n*\*one
\shl**{\dt*}\advance\n*\*one
\shl**{\dd*}\advance\n*\*two
\dd*-\dd*\ifnum\n*<\N*\repeat\n*\N*\shl**{0pt}}}

\def\wavebox#1{\setbox0\hbox{#1}\relax
\a*=\wd0\advance\a*14pt\b*=\ht0\advance\b*\dp0\advance\b*14pt\relax
\hbox{\kern9pt\relax
\rmov*(0pt,\ht0){\rmov*(-7pt,7pt){\wlin*\a*(1,0)[+]\wlin*\b*(0,-1)[-]}}\relax
\rmov*(\wd0,-\dp0){\rmov*(7pt,-7pt){\wlin*\a*(-1,0)[+]\wlin*\b*(0,1)[-]}}\relax
\box0\kern9pt}}

\def\rectangle#1(#2,#3){\relax
\lin#1(#2,0)\lin#1(0,#3)\mov#1(0,#3){\lin#1(#2,0)}\mov#1(#2,0){\lin#1(0,#3)}}

\def\dashrectangle#1(#2,#3){\dashlin#1(#2,0)\dashlin#1(0,#3)\relax
\mov#1(0,#3){\dashlin#1(#2,0)}\mov#1(#2,0){\dashlin#1(0,#3)}}

\def\waverectangle#1(#2,#3){\L*=#1\Lengthunit\a*=#2\L*\b*=#3\L*
\ifdim\a*<0pt\a*-\a*\def\x*{-1}\else\def\x*{1}\fi
\ifdim\b*<0pt\b*-\b*\def\y*{-1}\else\def\y*{1}\fi
\wlin*\a*(\x*,0)[-]\wlin*\b*(0,\y*)[+]\relax
\mov#1(0,#3){\wlin*\a*(\x*,0)[+]}\mov#1(#2,0){\wlin*\b*(0,\y*)[-]}}

\def\calcparab*{\ifnum\n*>\m*\k*\N*\advance\k*-\n*\else\k*\n*\fi
\a*=\the\k* sp\a*=10\a*\b*\dm*\advance\b*-\a*\k*\b*
\a*=\the\*ths\b*\divide\a*\l*\multiply\a*\k*
\divide\a*\l*\k*\*ths\r*\a*\advance\k*-\r*\dt*=\the\k*\L*}

\def\arcto#1(#2,#3)[#4]{\rlap{\toks0={#2}\toks1={#3}\calcnum*#1(#2,#3)\relax
\dm*=135sp\dm*=#1\dm*\d**=#1\Lengthunit\ifdim\dm*<0pt\dm*-\dm*\fi
\multiply\dm*\r*\a*=.3\dm*\a*=#4\a*\ifdim\a*<0pt\a*-\a*\fi
\advance\dm*\a*\N*\dm*\divide\N*10000\relax
\divide\N*2\multiply\N*2\advance\N*\*one
\L*=-.25\d**\L*=#4\L*\divide\d**\N*\divide\L*\*ths
\m*\N*\divide\m*2\dm*=\the\m*5sp\l*\dm*\sm*\n*\*one\loop
\calcparab*\shl**{-\dt*}\advance\n*1\ifnum\n*<\N*\repeat}}

\def\arrarcto#1(#2,#3)[#4]{\L*=#1\Lengthunit\L*=.54\L*
\arcto#1(#2,#3)[#4]\rmov*(#2\L*,#3\L*){\d*=.457\L*\d*=#4\d*\d**-\d*
\rmov*(#3\d**,#2\d*){\arrow.02(#2,#3)}}}

\def\dasharcto#1(#2,#3)[#4]{\rlap{\toks0={#2}\toks1={#3}\relax
\calcnum*#1(#2,#3)\dm*=\the\N*5sp\a*=.3\dm*\a*=#4\a*\ifdim\a*<0pt\a*-\a*\fi
\advance\dm*\a*\N*\dm*
\divide\N*20\multiply\N*2\advance\N*1\d**=#1\Lengthunit
\L*=-.25\d**\L*=#4\L*\divide\d**\N*\divide\L*\*ths
\m*\N*\divide\m*2\dm*=\the\m*5sp\l*\dm*
\sm*\n*\*one\loop\calcparab*
\shl**{-\dt*}\advance\n*1\ifnum\n*>\N*\else\calcparab*
\sh*(#2,#3){\xL*=#3\dt* \yL*=#2\dt*
\rx* \the\cos*\xL* \tmp* \the\sin*\yL* \advance\rx*\tmp*
\ry* \the\cos*\yL* \tmp* \the\sin*\xL* \advance\ry*-\tmp*
\kern\rx*\lower\ry*\hbox{\sm*}}\fi
\advance\n*1\ifnum\n*<\N*\repeat}}

\def\*shl*#1{\c*=\the\n*\d**\advance\c*#1\a**\d*\dt*\advance\d*#1\b**
\a*=\the\toks0\c*\b*=\the\toks1\d*\advance\a*-\b*
\b*=\the\toks1\c*\d*=\the\toks0\d*\advance\b*\d*
\rx* \the\cos*\a* \tmp* \the\sin*\b* \advance\rx*-\tmp*
\ry* \the\cos*\b* \tmp* \the\sin*\a* \advance\ry*\tmp*
\raise\ry*\rlap{\kern\rx*\unhcopy\spl*}}

\def\calcnormal*#1{\b**=10000sp\a**\b**\k*\n*\advance\k*-\m*
\multiply\a**\k*\divide\a**\m*\a**=#1\a**\ifdim\a**<0pt\a**-\a**\fi
\ifdim\a**>\b**\d*=.96\a**\advance\d*.4\b**
\else\d*=.96\b**\advance\d*.4\a**\fi
\d*=.01\d*\r*\d*\divide\a**\r*\divide\b**\r*
\ifnum\k*<0\a**-\a**\fi\d*=#1\d*\ifdim\d*<0pt\b**-\b**\fi
\k*\a**\a**=\the\k*\dd*\k*\b**\b**=\the\k*\dd*}

\def\wavearcto#1(#2,#3)[#4]{\rlap{\toks0={#2}\toks1={#3}\relax
\calcnum*#1(#2,#3)\c*=\the\N*5sp\a*=.4\c*\a*=#4\a*\ifdim\a*<0pt\a*-\a*\fi
\advance\c*\a*\N*\c*\divide\N*20\multiply\N*2\advance\N*-1\multiply\N*4\relax
\d**=#1\Lengthunit\dd*=.012\d**
\divide\dd*\*ths \multiply\dd*\magnitude
\ifdim\d**<0pt\d**-\d**\fi\L*=.25\d**
\divide\d**\N*\divide\dd*\N*\L*=#4\L*\divide\L*\*ths
\m*\N*\divide\m*2\dm*=\the\m*0sp\l*\dm*
\sm*\n*\*one\loop\calcnormal*{#4}\calcparab*
\*shl*{1}\advance\n*\*one\calcparab*
\*shl*{1.3}\advance\n*\*one\calcparab*
\*shl*{1}\advance\n*2\dd*-\dd*\ifnum\n*<\N*\repeat\n*\N*\shl**{0pt}}}

\def\triangarcto#1(#2,#3)[#4]{\rlap{\toks0={#2}\toks1={#3}\relax
\calcnum*#1(#2,#3)\c*=\the\N*5sp\a*=.4\c*\a*=#4\a*\ifdim\a*<0pt\a*-\a*\fi
\advance\c*\a*\N*\c*\divide\N*20\multiply\N*2\advance\N*-1\multiply\N*2\relax
\d**=#1\Lengthunit\dd*=.012\d**
\divide\dd*\*ths \multiply\dd*\magnitude
\ifdim\d**<0pt\d**-\d**\fi\L*=.25\d**
\divide\d**\N*\divide\dd*\N*\L*=#4\L*\divide\L*\*ths
\m*\N*\divide\m*2\dm*=\the\m*0sp\l*\dm*
\sm*\n*\*one\loop\calcnormal*{#4}\calcparab*
\*shl*{1}\advance\n*2\dd*-\dd*\ifnum\n*<\N*\repeat\n*\N*\shl**{0pt}}}

\def\hr*#1{\L*=\xscale\Lengthunit\ifnum
\angle**=0\clap{\vrule width#1\L* height.1pt}\else
\L*=#1\L*\L*=.5\L*\rmov*(-\L*,0pt){\sm*}\rmov*(\L*,0pt){\sl*}\fi}

\def\shade#1[#2]{\rlap{\Lengthunit=#1\Lengthunit
\special{em:linewidth .001pt}\relax
\mov(0,#2.05){\hr*{.994}}\mov(0,#2.1){\hr*{.980}}\relax
\mov(0,#2.15){\hr*{.953}}\mov(0,#2.2){\hr*{.916}}\relax
\mov(0,#2.25){\hr*{.867}}\mov(0,#2.3){\hr*{.798}}\relax
\mov(0,#2.35){\hr*{.715}}\mov(0,#2.4){\hr*{.603}}\relax
\mov(0,#2.45){\hr*{.435}}\special{em:linewidth \the\linwid*}}}

\def\dshade#1[#2]{\rlap{\special{em:linewidth .001pt}\relax
\Lengthunit=#1\Lengthunit\if#2-\def\t*{+}\else\def\t*{-}\fi
\mov(0,\t*.025){\relax
\mov(0,#2.05){\hr*{.995}}\mov(0,#2.1){\hr*{.988}}\relax
\mov(0,#2.15){\hr*{.969}}\mov(0,#2.2){\hr*{.937}}\relax
\mov(0,#2.25){\hr*{.893}}\mov(0,#2.3){\hr*{.836}}\relax
\mov(0,#2.35){\hr*{.760}}\mov(0,#2.4){\hr*{.662}}\relax
\mov(0,#2.45){\hr*{.531}}\mov(0,#2.5){\hr*{.320}}\relax
\special{em:linewidth \the\linwid*}}}}

\def\vdot{\rlap{\kern-1.9pt\lower1.8pt\hbox{$\scriptstyle\bullet$}}}
\def\vtimes{\rlap{\kern-3pt\lower1.8pt\hbox{$\scriptstyle\times$}}}
\def\vDot{\rlap{\kern-2.3pt\lower2.7pt\hbox{$\bullet$}}}
\def\vTimes{\rlap{\kern-3.6pt\lower2.4pt\hbox{$\times$}}}

\def\arc(#1)[#2,#3]{{\k*=#2\l*=#3\m*=\l*
\advance\m*-6\ifnum\k*>\l*\relax\else
{\rotate(#2)\mov(#1,0){\sm*}}\loop
\ifnum\k*<\m*\advance\k*5{\rotate(\k*)\mov(#1,0){\sl*}}\repeat
{\rotate(#3)\mov(#1,0){\sl*}}\fi}}

\def\dasharc(#1)[#2,#3]{{\k**=#2\n*=#3\advance\n*-1\advance\n*-\k**
\L*=1000sp\L*#1\L* \multiply\L*\n* \multiply\L*\Nhalfperiods
\divide\L*57\N*\L* \divide\N*2000\ifnum\N*=0\N*1\fi
\r*\n*	\divide\r*\N* \ifnum\r*<2\r*2\fi
\m**\r* \divide\m**2 \l**\r* \advance\l**-\m** \N*\n* \divide\N*\r*
\k**\r* \multiply\k**\N* \dn*\n* \advance\dn*-\k** \divide\dn*2\advance\dn*\*one
\r*\l** \divide\r*2\advance\dn*\r* \advance\N*-2\k**#2\relax
\ifnum\l**<6{\rotate(#2)\mov(#1,0){\sm*}}\advance\k**\dn*
{\rotate(\k**)\mov(#1,0){\sl*}}\advance\k**\m**
{\rotate(\k**)\mov(#1,0){\sm*}}\loop
\advance\k**\l**{\rotate(\k**)\mov(#1,0){\sl*}}\advance\k**\m**
{\rotate(\k**)\mov(#1,0){\sm*}}\advance\N*-1\ifnum\N*>0\repeat
{\rotate(#3)\mov(#1,0){\sl*}}\else\advance\k**\dn*
\arc(#1)[#2,\k**]\loop\advance\k**\m** \r*\k**
\advance\k**\l** {\arc(#1)[\r*,\k**]}\relax
\advance\N*-1\ifnum\N*>0\repeat
\advance\k**\m**\arc(#1)[\k**,#3]\fi}}

\def\triangarc#1(#2)[#3,#4]{{\k**=#3\n*=#4\advance\n*-\k**
\L*=1000sp\L*#2\L* \multiply\L*\n* \multiply\L*\Nhalfperiods
\divide\L*57\N*\L* \divide\N*1000\ifnum\N*=0\N*1\fi
\d**=#2\Lengthunit \d*\d** \divide\d*57\multiply\d*\n*
\r*\n*	\divide\r*\N* \ifnum\r*<2\r*2\fi
\m**\r* \divide\m**2 \l**\r* \advance\l**-\m** \N*\n* \divide\N*\r*
\dt*\d* \divide\dt*\N* \dt*.5\dt* \dt*#1\dt*
\divide\dt*1000\multiply\dt*\magnitude
\k**\r* \multiply\k**\N* \dn*\n* \advance\dn*-\k** \divide\dn*2\relax
\r*\l** \divide\r*2\advance\dn*\r* \advance\N*-1\k**#3\relax
{\rotate(#3)\mov(#2,0){\sm*}}\advance\k**\dn*
{\rotate(\k**)\mov(#2,0){\sl*}}\advance\k**-\m**\advance\l**\m**\loop\dt*-\dt*
\d*\d** \advance\d*\dt*
\advance\k**\l**{\rotate(\k**)\rmov*(\d*,0pt){\sl*}}%
\advance\N*-1\ifnum\N*>0\repeat\advance\k**\m**
{\rotate(\k**)\mov(#2,0){\sl*}}{\rotate(#4)\mov(#2,0){\sl*}}}}

\def\wavearc#1(#2)[#3,#4]{{\k**=#3\n*=#4\advance\n*-\k**
\L*=4000sp\L*#2\L* \multiply\L*\n* \multiply\L*\Nhalfperiods
\divide\L*57\N*\L* \divide\N*1000\ifnum\N*=0\N*1\fi
\d**=#2\Lengthunit \d*\d** \divide\d*57\multiply\d*\n*
\r*\n*	\divide\r*\N* \ifnum\r*=0\r*1\fi
\m**\r* \divide\m**2 \l**\r* \advance\l**-\m** \N*\n* \divide\N*\r*
\dt*\d* \divide\dt*\N* \dt*.7\dt* \dt*#1\dt*
\divide\dt*1000\multiply\dt*\magnitude
\k**\r* \multiply\k**\N* \dn*\n* \advance\dn*-\k** \divide\dn*2\relax
\divide\N*4\advance\N*-1\k**#3\relax
{\rotate(#3)\mov(#2,0){\sm*}}\advance\k**\dn*
{\rotate(\k**)\mov(#2,0){\sl*}}\advance\k**-\m**\advance\l**\m**\loop\dt*-\dt*
\d*\d** \advance\d*\dt* \dd*\d** \advance\dd*1.3\dt*
\advance\k**\r*{\rotate(\k**)\rmov*(\d*,0pt){\sl*}}\relax
\advance\k**\r*{\rotate(\k**)\rmov*(\dd*,0pt){\sl*}}\relax
\advance\k**\r*{\rotate(\k**)\rmov*(\d*,0pt){\sl*}}\relax
\advance\k**\r*
\advance\N*-1\ifnum\N*>0\repeat\advance\k**\m**
{\rotate(\k**)\mov(#2,0){\sl*}}{\rotate(#4)\mov(#2,0){\sl*}}}}

\def\gmov*#1(#2,#3)#4{\rlap{\L*=#1\Lengthunit
\xL*=#2\L* \yL*=#3\L*
\rx* \gcos*\xL* \tmp* \gsin*\yL* \advance\rx*-\tmp*
\ry* \gcos*\yL* \tmp* \gsin*\xL* \advance\ry*\tmp*
\rx*=\xscale\rx* \ry*=\yscale\ry*
\xL* \the\cos*\rx* \tmp* \the\sin*\ry* \advance\xL*-\tmp*
\yL* \the\cos*\ry* \tmp* \the\sin*\rx* \advance\yL*\tmp*
\kern\xL*\raise\yL*\hbox{#4}}}

\def\rgmov*(#1,#2)#3{\rlap{\xL*#1\yL*#2\relax
\rx* \gcos*\xL* \tmp* \gsin*\yL* \advance\rx*-\tmp*
\ry* \gcos*\yL* \tmp* \gsin*\xL* \advance\ry*\tmp*
\rx*=\xscale\rx* \ry*=\yscale\ry*
\xL* \the\cos*\rx* \tmp* \the\sin*\ry* \advance\xL*-\tmp*
\yL* \the\cos*\ry* \tmp* \the\sin*\rx* \advance\yL*\tmp*
\kern\xL*\raise\yL*\hbox{#3}}}

\def\Earc(#1)[#2,#3][#4,#5]{{\k*=#2\l*=#3\m*=\l*
\advance\m*-6\ifnum\k*>\l*\relax\else\def\xscale{#4}\def\yscale{#5}\relax
{\angle**0\rotate(#2)}\gmov*(#1,0){\sm*}\loop
\ifnum\k*<\m*\advance\k*5\relax
{\angle**0\rotate(\k*)}\gmov*(#1,0){\sl*}\repeat
{\angle**0\rotate(#3)}\gmov*(#1,0){\sl*}\relax
\def\xscale{1}\def\yscale{1}\fi}}

\def\dashEarc(#1)[#2,#3][#4,#5]{{\k**=#2\n*=#3\advance\n*-1\advance\n*-\k**
\L*=1000sp\L*#1\L* \multiply\L*\n* \multiply\L*\Nhalfperiods
\divide\L*57\N*\L* \divide\N*2000\ifnum\N*=0\N*1\fi
\r*\n*	\divide\r*\N* \ifnum\r*<2\r*2\fi
\m**\r* \divide\m**2 \l**\r* \advance\l**-\m** \N*\n* \divide\N*\r*
\k**\r*\multiply\k**\N* \dn*\n* \advance\dn*-\k** \divide\dn*2\advance\dn*\*one
\r*\l** \divide\r*2\advance\dn*\r* \advance\N*-2\k**#2\relax
\ifnum\l**<6\def\xscale{#4}\def\yscale{#5}\relax
{\angle**0\rotate(#2)}\gmov*(#1,0){\sm*}\advance\k**\dn*
{\angle**0\rotate(\k**)}\gmov*(#1,0){\sl*}\advance\k**\m**
{\angle**0\rotate(\k**)}\gmov*(#1,0){\sm*}\loop
\advance\k**\l**{\angle**0\rotate(\k**)}\gmov*(#1,0){\sl*}\advance\k**\m**
{\angle**0\rotate(\k**)}\gmov*(#1,0){\sm*}\advance\N*-1\ifnum\N*>0\repeat
{\angle**0\rotate(#3)}\gmov*(#1,0){\sl*}\def\xscale{1}\def\yscale{1}\else
\advance\k**\dn* \Earc(#1)[#2,\k**][#4,#5]\loop\advance\k**\m** \r*\k**
\advance\k**\l** {\Earc(#1)[\r*,\k**][#4,#5]}\relax
\advance\N*-1\ifnum\N*>0\repeat
\advance\k**\m**\Earc(#1)[\k**,#3][#4,#5]\fi}}

\def\triangEarc#1(#2)[#3,#4][#5,#6]{{\k**=#3\n*=#4\advance\n*-\k**
\L*=1000sp\L*#2\L* \multiply\L*\n* \multiply\L*\Nhalfperiods
\divide\L*57\N*\L* \divide\N*1000\ifnum\N*=0\N*1\fi
\d**=#2\Lengthunit \d*\d** \divide\d*57\multiply\d*\n*
\r*\n*	\divide\r*\N* \ifnum\r*<2\r*2\fi
\m**\r* \divide\m**2 \l**\r* \advance\l**-\m** \N*\n* \divide\N*\r*
\dt*\d* \divide\dt*\N* \dt*.5\dt* \dt*#1\dt*
\divide\dt*1000\multiply\dt*\magnitude
\k**\r* \multiply\k**\N* \dn*\n* \advance\dn*-\k** \divide\dn*2\relax
\r*\l** \divide\r*2\advance\dn*\r* \advance\N*-1\k**#3\relax
\def\xscale{#5}\def\yscale{#6}\relax
{\angle**0\rotate(#3)}\gmov*(#2,0){\sm*}\advance\k**\dn*
{\angle**0\rotate(\k**)}\gmov*(#2,0){\sl*}\advance\k**-\m**
\advance\l**\m**\loop\dt*-\dt* \d*\d** \advance\d*\dt*
\advance\k**\l**{\angle**0\rotate(\k**)}\rgmov*(\d*,0pt){\sl*}\relax
\advance\N*-1\ifnum\N*>0\repeat\advance\k**\m**
{\angle**0\rotate(\k**)}\gmov*(#2,0){\sl*}\relax
{\angle**0\rotate(#4)}\gmov*(#2,0){\sl*}\def\xscale{1}\def\yscale{1}}}

\def\waveEarc#1(#2)[#3,#4][#5,#6]{{\k**=#3\n*=#4\advance\n*-\k**
\L*=4000sp\L*#2\L* \multiply\L*\n* \multiply\L*\Nhalfperiods
\divide\L*57\N*\L* \divide\N*1000\ifnum\N*=0\N*1\fi
\d**=#2\Lengthunit \d*\d** \divide\d*57\multiply\d*\n*
\r*\n*	\divide\r*\N* \ifnum\r*=0\r*1\fi
\m**\r* \divide\m**2 \l**\r* \advance\l**-\m** \N*\n* \divide\N*\r*
\dt*\d* \divide\dt*\N* \dt*.7\dt* \dt*#1\dt*
\divide\dt*1000\multiply\dt*\magnitude
\k**\r* \multiply\k**\N* \dn*\n* \advance\dn*-\k** \divide\dn*2\relax
\divide\N*4\advance\N*-1\k**#3\def\xscale{#5}\def\yscale{#6}\relax
{\angle**0\rotate(#3)}\gmov*(#2,0){\sm*}\advance\k**\dn*
{\angle**0\rotate(\k**)}\gmov*(#2,0){\sl*}\advance\k**-\m**
\advance\l**\m**\loop\dt*-\dt*
\d*\d** \advance\d*\dt* \dd*\d** \advance\dd*1.3\dt*
\advance\k**\r*{\angle**0\rotate(\k**)}\rgmov*(\d*,0pt){\sl*}\relax
\advance\k**\r*{\angle**0\rotate(\k**)}\rgmov*(\dd*,0pt){\sl*}\relax
\advance\k**\r*{\angle**0\rotate(\k**)}\rgmov*(\d*,0pt){\sl*}\relax
\advance\k**\r*
\advance\N*-1\ifnum\N*>0\repeat\advance\k**\m**
{\angle**0\rotate(\k**)}\gmov*(#2,0){\sl*}\relax
{\angle**0\rotate(#4)}\gmov*(#2,0){\sl*}\def\xscale{1}\def\yscale{1}}}

\newcount\CatcodeOfAtSign
\CatcodeOfAtSign=\the\catcode`\@
\catcode`\@=11
\def\@arc#1[#2][#3]{\rlap{\Lengthunit=#1\Lengthunit
\sm*\l*arc(#2.1914,#3.0381)[#2][#3]\relax
\mov(#2.1914,#3.0381){\l*arc(#2.1622,#3.1084)[#2][#3]}\relax
\mov(#2.3536,#3.1465){\l*arc(#2.1084,#3.1622)[#2][#3]}\relax
\mov(#2.4619,#3.3086){\l*arc(#2.0381,#3.1914)[#2][#3]}}}

\def\dash@arc#1[#2][#3]{\rlap{\Lengthunit=#1\Lengthunit
\d*arc(#2.1914,#3.0381)[#2][#3]\relax
\mov(#2.1914,#3.0381){\d*arc(#2.1622,#3.1084)[#2][#3]}\relax
\mov(#2.3536,#3.1465){\d*arc(#2.1084,#3.1622)[#2][#3]}\relax
\mov(#2.4619,#3.3086){\d*arc(#2.0381,#3.1914)[#2][#3]}}}

\def\wave@arc#1[#2][#3]{\rlap{\Lengthunit=#1\Lengthunit
\w*lin(#2.1914,#3.0381)\relax
\mov(#2.1914,#3.0381){\w*lin(#2.1622,#3.1084)}\relax
\mov(#2.3536,#3.1465){\w*lin(#2.1084,#3.1622)}\relax
\mov(#2.4619,#3.3086){\w*lin(#2.0381,#3.1914)}}}

\def\bezier#1(#2,#3)(#4,#5)(#6,#7){\N*#1\l*\N* \advance\l*\*one
\d* #4\Lengthunit \advance\d* -#2\Lengthunit \multiply\d* \*two
\b* #6\Lengthunit \advance\b* -#2\Lengthunit
\advance\b*-\d* \divide\b*\N*
\d** #5\Lengthunit \advance\d** -#3\Lengthunit \multiply\d** \*two
\b** #7\Lengthunit \advance\b** -#3\Lengthunit
\advance\b** -\d** \divide\b**\N*
\mov(#2,#3){\sm*{\loop\ifnum\m*<\l*
\a*\m*\b* \advance\a*\d* \divide\a*\N* \multiply\a*\m*
\a**\m*\b** \advance\a**\d** \divide\a**\N* \multiply\a**\m*
\rmov*(\a*,\a**){\unhcopy\spl*}\advance\m*\*one\repeat}}}

\catcode`\*=12

\newcount\n@ast
\def\n@ast@#1{\n@ast0\relax\get@ast@#1\end}
\def\get@ast@#1{\ifx#1\end\let\next\relax\else
\ifx#1*\advance\n@ast1\fi\let\next\get@ast@\fi\next}

\newif\if@up \newif\if@dwn
\def\up@down@#1{\@upfalse\@dwnfalse
\if#1u\@uptrue\fi\if#1U\@uptrue\fi\if#1+\@uptrue\fi
\if#1d\@dwntrue\fi\if#1D\@dwntrue\fi\if#1-\@dwntrue\fi}

\def\halfcirc#1(#2)[#3]{{\Lengthunit=#2\Lengthunit\up@down@{#3}\relax
\if@up\mov(0,.5){\@arc[-][-]\@arc[+][-]}\fi
\if@dwn\mov(0,-.5){\@arc[-][+]\@arc[+][+]}\fi
\def\lft{\mov(0,.5){\@arc[-][-]}\mov(0,-.5){\@arc[-][+]}}\relax
\def\rght{\mov(0,.5){\@arc[+][-]}\mov(0,-.5){\@arc[+][+]}}\relax
\if#3l\lft\fi\if#3L\lft\fi\if#3r\rght\fi\if#3R\rght\fi
\n@ast@{#1}\relax
\ifnum\n@ast>0\if@up\shade[+]\fi\if@dwn\shade[-]\fi\fi
\ifnum\n@ast>1\if@up\dshade[+]\fi\if@dwn\dshade[-]\fi\fi}}

\def\halfdashcirc(#1)[#2]{{\Lengthunit=#1\Lengthunit\up@down@{#2}\relax
\if@up\mov(0,.5){\dash@arc[-][-]\dash@arc[+][-]}\fi
\if@dwn\mov(0,-.5){\dash@arc[-][+]\dash@arc[+][+]}\fi
\def\lft{\mov(0,.5){\dash@arc[-][-]}\mov(0,-.5){\dash@arc[-][+]}}\relax
\def\rght{\mov(0,.5){\dash@arc[+][-]}\mov(0,-.5){\dash@arc[+][+]}}\relax
\if#2l\lft\fi\if#2L\lft\fi\if#2r\rght\fi\if#2R\rght\fi}}

\def\halfwavecirc(#1)[#2]{{\Lengthunit=#1\Lengthunit\up@down@{#2}\relax
\if@up\mov(0,.5){\wave@arc[-][-]\wave@arc[+][-]}\fi
\if@dwn\mov(0,-.5){\wave@arc[-][+]\wave@arc[+][+]}\fi
\def\lft{\mov(0,.5){\wave@arc[-][-]}\mov(0,-.5){\wave@arc[-][+]}}\relax
\def\rght{\mov(0,.5){\wave@arc[+][-]}\mov(0,-.5){\wave@arc[+][+]}}\relax
\if#2l\lft\fi\if#2L\lft\fi\if#2r\rght\fi\if#2R\rght\fi}}

\catcode`\*=11

\def\Circle#1(#2){\halfcirc#1(#2)[u]\halfcirc#1(#2)[d]\n@ast@{#1}\relax
\ifnum\n@ast>0\L*=\xscale\Lengthunit
\ifnum\angle**=0\clap{\vrule width#2\L* height.1pt}\else
\L*=#2\L*\L*=.5\L*\special{em:linewidth .001pt}\relax
\rmov*(-\L*,0pt){\sm*}\rmov*(\L*,0pt){\sl*}\relax
\special{em:linewidth \the\linwid*}\fi\fi}

\catcode`\*=12

\def\wavecirc(#1){\halfwavecirc(#1)[u]\halfwavecirc(#1)[d]}

\def\dashcirc(#1){\halfdashcirc(#1)[u]\halfdashcirc(#1)[d]}

\def\xscale{1}
\def\yscale{1}

\def\Ellipse#1(#2)[#3,#4]{\def\xscale{#3}\def\yscale{#4}\relax
\Circle#1(#2)\def\xscale{1}\def\yscale{1}}

\def\dashEllipse(#1)[#2,#3]{\def\xscale{#2}\def\yscale{#3}\relax
\dashcirc(#1)\def\xscale{1}\def\yscale{1}}

\def\waveEllipse(#1)[#2,#3]{\def\xscale{#2}\def\yscale{#3}\relax
\wavecirc(#1)\def\xscale{1}\def\yscale{1}}

\def\halfEllipse#1(#2)[#3][#4,#5]{\def\xscale{#4}\def\yscale{#5}\relax
\halfcirc#1(#2)[#3]\def\xscale{1}\def\yscale{1}}

\def\halfdashEllipse(#1)[#2][#3,#4]{\def\xscale{#3}\def\yscale{#4}\relax
\halfdashcirc(#1)[#2]\def\xscale{1}\def\yscale{1}}

\def\halfwaveEllipse(#1)[#2][#3,#4]{\def\xscale{#3}\def\yscale{#4}\relax
\halfwavecirc(#1)[#2]\def\xscale{1}\def\yscale{1}}

\catcode`\@=\the\CatcodeOfAtSign

\begin{center}
{\bf CORRECTIONS OF ORDER $(Z\alpha)^6 \frac{m_e^2}{m_\mu}$\\
IN THE MUONIUM FINE STRUCTURE}\footnote{Talk presented at the
International Workshop "Hadronic Atoms and Positronium in the
Standard Model" Dubna, 26-31 May 1998}\\

\vspace{4mm}

R.N.~Faustov \\Scientific Council "Cybernetics" RAS\\
117333, Moscow, Vavilov, 40, Russia,\\
A.P.~Martynenko\\ Department of Theoretical Physics, Samara State University,\\
443011, Samara, Pavlov, 1, Russia
\end{center}

\begin{abstract}
In the framework of the quasipotential method we calculate the
contributions of the kind $(Z\alpha)^6m_e^2/m_\mu$
to the energy spectrum of the muonium $n^3S_1$ states. Numerical value of
obtained correction for $2^3S_1\div 1^3S_1$ muonium fine structure
interval is equal to 0.19 MHz.
\end{abstract}

\newpage

The investigation of muonium and positronium fine structure
represents one of the basic tests of quantum electrodynamics, which is
sensitive to radiative corrections of higher order on $\alpha$ \cite{KS}.
Many papers \cite{Ful,Sal,GY,GE} are devoted to the calculation of different
contributions to the fine structure  of hydrogen-like atom energy levels.
The interest to this problem remains unchanged \cite{PG,Y,EG}. The progress,
achieved in the last years during the calculation of logarithmic
contributions of order $\alpha^6 \ln\alpha$ in the positronium fine
structure intervals $(2^3S_1\div 1^3S_1,
2^3S_1\div 2^3P_J)$ \cite{Kh, Fell, FK}, doesn't abolish the necessity of
calculation of higher order corrections $O(\alpha^6)$ \cite{F}.
As discussed in the paper \cite{E}, there is some difference in the
calculation of the corrections $(Z\alpha)^6m_1^2/m_2$, obtained in \cite{EG,Y}.
The development of experimental methods, based on Doppler-free two-photon
spectroscopy, allows the "large" structure intervals to be measured for
the muonium and the positronium \cite{Chu, Chu1, Fee}.
\begin{equation}
\Delta E^{exp.}_{Ps}(2^3S_1\div 1^3S_1)=\left\{{1233607218,9\pm10,7~MHz}
\atop{1233607216,4\pm 3,2~MHz}\right.
\end{equation}
\begin{equation}
\Delta E^{exp.}_{Mu}(2^3S_1\div 1^3S_1)=2455527936\pm 120\pm 140~MHz.
\end{equation}

The frequency of Doppler-free two-photon transition $1S\div 2S$ in the
hydrogen atom as well as hyperfine splitting of the ground state of
hydrogen atom represents the quantity, which was measured with high accuracy
\cite{F}. The increase of the experimental accuracy of the muonium fine structure
interval measurements (just as for positronium), planned in the near future, makes very actual
the calculation of radiative corrections of higher order on
$\alpha$. In this paper we have performed studies of the recoil contributions
of order $O(\alpha^6)$ in the muonium fine structure. The contribution of such order
to the muonium hyperfine structure was obtained in \cite{BYG}. There exist many
approaches for description of relativistic energy spectrums of two-particle
bound states in quantum electrodynamics \cite{Sal,GY,IBK,CL,RT}. They all differ in
organization of the calculation: the bound state equation for two-particle
system, the construction of the interaction operator, the degree of
complication in the calculation of the definite order corrections in the
energy levels. But all such methods give equivalent results in the fixed
order of perturbation theory on the small parameter $\alpha$ and $m_1/m_2$.
Our calculations
are based on the Schrodinger-type local quasipotential equation \cite{FM}
\begin{equation}
\left(\frac{b^2}{2\mu_R}-\frac{\vec p^2}{2\mu_R}\right)\psi_M(\vec p)=
\int\frac{d\vec q}{(2\pi)^3}V(\vec p,\vec q,M)\psi_M(\vec q),
\end{equation}
where $b^2=E^2_1-m^2_1=E^2_2-m^2_2,~\mu_R=E_1E_2/M$ is relativistic reduced mass,
$M=E_1+E_2$ is the bound state mass, $m_1, m_2$ are the masses of the electron and
the muon. As an initial approximation of quasipotential $V(\vec p,\vec q,M)$
for the bound state system $(e^-\mu^+)$ we choose the ordinary Coulomb
potential. On the basis of equation (3) in \cite{FM1} we have obtained some
relativistic corrections $m\alpha^6$ in the positronium fine structure from
the one-photon, two-photon interaction and the second order perturbation
theory. First of all, the contribution of order $(Z\alpha)^6m_1/m_2$
\begin{equation}
\Delta B_1=\frac{5m_1^2(Z\alpha)^6}{2m_2n^6}
\end{equation}
appears from the condition of quantization of the energy levels for Coulomb
interaction
\begin{equation}
\frac{b^2}{\mu_R^2}=-\frac{\alpha^2}{n^2},
\end{equation}
which we have transformed for binding energy B.

\section{One-photon interaction contribution to the fine structure}

The basic contribution to the energy spectrum of two-particle bound state
($\mu^+ e^-$) is determined by one-photon interaction. The construction of
one-photon quasipotential in the system of two spinor particles was done in 
\cite{FM1,FM2}. It is useful to note, that procedure of quasipotential
derivation may be essentially simplified in this case, just as for two-photon
and three-photon exchange diagrams by introducing the relativistic operator
of projection onto $^3S_1$ muonium state:
\begin{equation}
\hat\Pi=\frac{1}{2\sqrt{2}}\frac{(\hat p_1+m_1)(1+\gamma_0)\hat\varepsilon(-\hat p_2+m_2)}
{\sqrt{\varepsilon_1+m_1}\sqrt{\varepsilon_2+m_2}}
\end{equation}
(where $p_1$, $p_2$ are four-momenta of electron and muon in the initial state,
$\varepsilon^\mu$ is the muonium polarization vector. Doing the decomposition
of all relativistic factors over two small parameters:
$|\vec p|/m_{1,2}$ ($|\vec p|\sim Z\alpha$ is the momentum of relative motion)
and $m_1/m_2$ in order to extract the contribution of six order over $\alpha$
and the first order over $m_1/m_2$ we can present the interaction operator in
coordinate space in the following manner:
\begin{equation}
V_1(r)=-\frac{Z\alpha}{r}-\frac{\mu_R(Z\alpha)^2}{2m_1^2r^2}\left(1+\frac{2m_1}{m_2}\right)-
\frac{Z\alpha}{4m_1^2r^3}(\vec r\nabla)\left(1+\frac{4m_1}{m_2}\right)-
\end{equation}
\begin{displaymath}
-\frac{\pi Z\alpha}{3m_1m_2}\delta(\vec r)-\frac{Z\alpha b^2}{m_1m_2r}=
V_c(r)+\Delta V_1(r).
\end{displaymath}

It is necessary to point out, that the part of quasipotential (7) $\Delta V_1$
taking together with quantization condition (5) reproduce correctly the well-known
energy spectrum of muonium S-wave states with the accuracy $O(\alpha^4)$ \cite{F,RT}.
The terms of $\Delta V_1$ in our approach give also the corrections
$O(\alpha^6)$ in the energy levels. The reason is that the relativistic
reduced mass $\mu_R$ and $b^2$ are dependant from $\alpha$. Averaging (7)
over Coulomb wave functions \cite{t4} and extracting necessary order
contributions, we have obtained:
\begin{equation}
\Delta B_2=-\frac{3m_1^2(Z\alpha)^6}{4m_2n^5}\left(5+\frac{2}{n}\right).
\end{equation}

The quasipotential of one-photon interaction contains also a number of other
terms, which lead to the energy corrections of order $O(Z\alpha)^6$. They may
be found  when constructing $V_{1\gamma}$ with the accuracy to terms
of fourth order over $|\vec p|/m_{1,2}$, $|\vec q|/m_{1,2}$ in the form
\begin{equation}
\Delta V_2(\vec p,\vec q,M)=-\frac{4\pi Z\alpha}{\vec k^2}\Bigl\{\frac{b^4}
{16m_1^4}\left(3-\frac{2m_1}{m_2}\right)+
+\frac{\vec p^4+\vec q^4}{96m_1^4}\left(3+\frac{m_1}{m_2}\right)-
\end{equation}
\begin{displaymath}
-\frac{(\vec p^2+\vec q^2)(\vec p\vec q)}{96m_1^4}\left(6+\frac{13m_1}{m_2}\right)-
\frac{(\vec p^2+\vec q^2)b^2}{96m_1^4}\left(3-\frac{m_1}{m_2}\right)-
\frac{(\vec p\vec q)b^2}{48m_1^4}\left(3+\frac{7m_1}{m_2}\right)\Bigr\}.
\end{displaymath}

Let consider calculation of $O(Z\alpha)^6$ corrections from $\Delta V_2$.
A series of $\Delta V_2$ terms will cause the divergent integrals in the
energy spectrum. The reason behind this divergence is that, in deriving (9),
we expanded all quantities determining the quasipotential in the relative
momenta, which are proportional to $\alpha$. A typical divergent integral with
respect to momenta has the form $\int \vec p^2d\vec p\psi_{nS}(\vec p)$.
The relativistic correction of order $\alpha^6$ in this case is determined by the
residue of the integrand at the pole of the wave function $\psi_{nS}(\vec p)$
\cite{IBK}. The result of this integral calculation for arbitrary principal
quantum number n has the form:
\begin{equation}
\int\frac{d\vec p}{(2\pi)^3}\frac{\vec p^2}{\mu_R^2}\psi_{nS}(\vec p)=
-\frac{[3+2(n-1)(n+1)]}{n^2}\alpha^2\psi_{nS}(\vec r=0).
\end{equation}
Taking into account (9) under averaging $\Delta V_2$ we have calculated the
relativistic corrections of necessary order for the energy levels with arbitrary
principal quantum number n:
\begin{equation}
\Delta B_3=\frac{m_1^2(Z\alpha)^6}{m_2}\Biggl(-\frac{139}{72n^3}+\frac{17}{12}
\frac{\ln 2}{n^3}+\frac{73}{72}\frac{1}{n^5}+\frac{43}{96}\frac{1}{n^6}
+\frac{17}{12}(-1)^n\frac{1}{n^3}[C+\psi(n)-1]\Biggr),
\end{equation}
where $\psi(z)=d\ln \Gamma(z)/dz$, C=0.5772156649... is the Euler constant.

\section{Second order of perturbation theory}

In the second order of perturbation theory, the correction to the muonium
energy spectrum is determined in our approach by the sum of two terms \cite{t2}:
\begin{equation}
\Delta B^{(2)}=<\psi^c_n|\Delta V_1|\psi^c_n><\psi^c_n|\frac{\partial
\Delta V_1}
{\partial B}|\psi^c_n>+
\sum_{k=1, k\neq n}^{\infty}\frac{<\psi^c_n|\Delta V_1|
\psi^c_k><\psi^c_k|\Delta V_1|\psi^c_k>}{B^c_n-B^c_k}.
\end{equation}
The quasipotential (9) explicitly depends on the bound state energy
B (the factors $b^2$ and $\mu_R$). Bearing in mind that to a precision adopted here, the
relation $\partial b^2/\partial B=2\mu$ holds, we obtain:
\begin{equation}
\Delta B_4=<\psi^c_n|\Delta V_1|\psi^c_n><\psi^c_n|\frac{\partial\Delta V_1}
{\partial B}|\psi^c_n>=\frac{m_1^2}{m_2}(Z\alpha)^6\frac{1}{n^5}.
\end{equation}
The sum over Coulomb states in (12) is defined by the reduced nonrelativistic
Green's function \cite{CL,Sch,Buch,Hostler,Vol,SGK,ZMP}, which has the
following partial expansion:
\begin{equation}
\bar G_n(\vec r,\vec r',B)=\sum_{l,m}\bar g_{nl}(r,r',B)Y_{lm}(\vec n)Y^{\ast}_{lm}(\vec n')
\end{equation}
The radial function $\bar g_{nl}(r,r',B_n)$ was obtained in \cite{ZMP} as an
expansion over Laguerre's polynomials. In the case of S-wave states we have for
it the next expression:
\begin{equation}
\bar g_{n0}(r,r',B_n)=-\frac{4Z\alpha\mu^2}{n}\Biggl[e^{-\frac{x+x'}{2}}\sum_{m=1,m\neq n}^\infty
\frac{L^1_{m-1}(x)L^1_{m-1}(x')}{m(m-n)}+
\end{equation}
\begin{displaymath}
+\frac{1}{n^2}\left(\frac{5}{2}+x\frac{\partial}{\partial x}+x'\frac{\partial}
{\partial x'}\right)e^{-\frac{x+x'}{2}}L^1_{n-1}(x)L^1_{n-1}(x')\Biggr],
\end{displaymath}
where $x=2\mu Z\alpha r/n$, $L^m_n$ are the ordinary Laguerre's polynomials,
which may be found by means the following relation:
\begin{equation}
L^m_n(x)=\frac{e^xx^{-m}}{n!}\left(\frac{d}{dx}\right)^n(e^{-x}x^{n+m}).
\end{equation}
The reduced Coulomb Green's function (RCGF) depends from two variables
r and r'. Due to  the appearance of the $\delta$ - like potential in (9) the
RCGF is needed for $\vec r=0$. The necessary expression for RCGF was obtained
by means of Hostler's representation for Coulomb Green's function in
\cite{IK}, after subtraction of pole-like term:
\begin{equation}
\bar G_n(\vec r,0,B_n)=-\frac{Z\alpha \mu^2}{n\pi x}e^{-\frac{x}{2}}\sum_
{s=0}^{n-1}\frac{(-x)^{n-s}}{s!}\frac{n!}{[(n-s)!]^2}
\end{equation}
\begin{displaymath}
\left\{(n-s)[\psi(n+1)-2\psi(n-s+1)-\frac{2(n-s)+3-x}{2n}+\ln x]+1\right\},
\end{displaymath}
where $\psi(z)=d\ln\Gamma(z)/dz$. Contrary to the paper \cite{IK} this formula
doesn't contain free two-particle Green's function $G^f(r)=-\mu_R
e^{-Z\alpha\mu_Rr}/2\pi r$, which controls the iteration part of quasipotential.
Its contribution to the energy spectrum will be find separately. Let consider,
for example, calculation of energy correction in the second order of
perturbation theory, depending from $\delta$-like part of the quasipotential
and the term of $\Delta V_1\sim 1/r^2$. This contribution can be represented
in the form:
\begin{equation}
\delta B=-\frac{\mu^5(Z\alpha)^6}{3m_1m_2n^4}\sum_{k=1}^n(-1)^k\frac{n!}{(n-k)!
(k!)^2}
\end{equation}
\begin{displaymath}
\int_0^\infty x^{k-1}e^{-x}L^1_{n-1}(x)dx\left\{k[\psi(n+1)-2\psi(k+1)-
\frac{2k+3-x}{2n}+\ln x]+1\right\}.
\end{displaymath}

The expression (18) contains the integrals of two types with the power and
the logarithmic functions correspondingly. Calculation of the first integral
over x variable leads to the result
\begin{equation}
I_1=\int_0^\infty x^{k-1}e^{-x}L^1_{n-1}(x)=\frac{(k-1)!\Gamma(n+1-k)}{(n-1)!
\Gamma(2-k)}.
\end{equation}
So, we see that the sum over k contains only one term with k=1. Second integral
of (18)
\begin{equation}
I_2=\int_0^\infty x^{k-1}\ln x e^{-x}L^1_{n-1}(x)=\frac{(2-k)_{n-1}\Gamma(k)}
{(n-1)!}[\psi(k)+\psi(2-k)-\psi(n+1-k)]
\end{equation}
gives rise the sum of the kind
\begin{equation}
\sum_{k=1}^n(-1)^k\frac{\psi(2-k)(2-k)_{n-1}}{(n-k)!k!}=-\frac{n-1}{n}+C,
\end{equation}
where $C=\lim_{n\rightarrow\infty}[-\ln n+\sum_{m=1}^n\frac{1}{m}]=0.57721566...$.
Taking into account that
\begin{equation}
\lim_{k\rightarrow n}\frac{\Gamma(n-k)}{\Gamma(1-k)}=(-1)^{n-1}(n-1)!,
\end{equation}
we have obtained conclusively respective correction (18):
\begin{equation}
\delta B=-\frac{m_1^2}{6m_2}(Z\alpha)^6\frac{1}{n^4}(4n+9).
\end{equation}

{\bf Table 1. Second order contributions of the perturbation theory, defined
by the RCGF and quasipotential (9) in the units $\frac{(Z\alpha)^6m_1^2}{m_2}$}\\[3mm]
\begin{tabular}{|c|c|c|c|c|}	 \hline
 &   &	 &   &	 \\
$\Delta V_1$ &	$-\frac{Z\alpha b^2}{m_1m_2 r}$   & $-\frac{(Z\alpha)^2\mu_R}
{2r^2m_1^2}(1+\frac{2m_1}{m_2})$  &  $-\frac{Z\alpha(\vec r\nabla)}{4m_1^2r^3}
(1+\frac{4m_1}{m_2})$ & $-\frac{\pi Z\alpha}{3m_1m_2}\delta(\vec r)$   \\
 &    &   &   &   \\	\hline
 &    &   &   &   \\
$-\frac{Z\alpha b^2}{m_1m_2r}$ & ---& $\frac{2-n}{n^5}$  & $\frac{2n^2-5n+1}{4n^6}$ &
---    \\
 &    &    &   &   \\	\hline
 &    &    &   &   \\
$-\frac{(Z\alpha)^2\mu_R}{2m_1^2 r^2}(1+\frac{2m_1}{m_2})$  & $\frac{2-n}{n^5}$ &
$\frac{2n+3}{2n^4}$ & $-\frac{n^2+3n-1}{4n^5}$ & $-\frac{(4n+9)}{6n^4} $   \\
 &    &    &   &    \\	 \hline
 &    &    &   &    \\
$-\frac{Z\alpha(\vec r\nabla)}{4m_1^2r^3}(1+\frac{4m_1}{m_2})$ &
$\frac{2n^2-5n+1}{4n^6}$ & $-\frac{n^2+3n-1}{4n^5}$ & $-\frac{n(n-1)(n+1)}{24n^6}$ &
$-\frac{(n^2-6n+8)}{12n^5}$  \\
  &   &    &   &    \\	 \hline
  &   &    &   &     \\
$-\frac{\pi Z\alpha}{3m_1m_2}\delta(\vec r)$ & --- & $-\frac{(4n+9)}{6n^4}$  &
$-\frac{(n^2-6n+8)}{12n^5}$ & ---  \\
 &    &    &   &    \\	\hline
\end{tabular}

\vspace{3mm}

Similar calculations may be carried out to find the contributions of the other
terms of quasipotential (9) in the second order of perturbation theory. The
respective results of such calculation represented in the Table 1. The full
correction, connected with reduced Coulomb Green's function without
considering $G^f$ in the energy spectrum is:
\begin{equation}
\Delta B_5=\left(-\frac{25}{24}-\frac{3}{n}-\frac{49}{24n^2}+\frac{3}{2n^3}\right)
\frac{m_1^2(Z\alpha)^6}{m_2n^3}
\end{equation}

Let consider now omitting in the RCGF contribution of free two-particle
propagator to the correction $\Delta B^{(2)}$. It is convenient to perform
necessary calculations in the momentum representation. Taking in mind that 
\begin{equation}
G^f(\vec p,\vec q,M)=\frac{(2\pi)^3\delta(\vec p-\vec q)}{\frac{b^2}{2\mu_R}-
\frac{\vec p^2}{2\mu_R}},
\end{equation}
and $\delta$-like term of quasipotential (9) already contains the muon mass in
the denominator, we may represent the iteration-type correction as:
\begin{equation}
\Delta B_6=\frac{2\mu(Z\alpha)^2\pi\psi_{nS}(0)}{3m_2m_1^3}\int\frac{d\vec p}
{(2\pi)^3}\psi_{nS}(\vec p)\left[\frac{\vec p\vec k}{\vec k^2}-
\frac{\vec p^2+W^2}{\vec k^2}\right]\frac{d\vec q}{(2\pi)^3(\vec q^2+W^2)},
\end{equation}
where $W^2=\mu_R^2(Z\alpha)^2/n^2$. The divergence of this integral identical
to that for (10). Using Feynman's parameterization and the substitution of
the type (10) to calculate (26), we obtain:
\begin{equation}
\Delta B_6=-\frac{m_1^2(Z\alpha)^6}{12m_2}\left(\frac{1}{n^3}-\frac{2\ln 2}{n^3}-
\frac{2}{n^4}-\frac{2}{n^3}(-1)^n[C+\psi(n)-1]\right).
\end{equation}

\section{Two-photon exchange interaction}

The amplitude of two-photon exchange interaction is represented on two
diagrams of Fig.1. The appropriate quasipotentials are:
\begin{equation}
V_{2\gamma}^{(a)}(\vec p,\vec q)=\frac{i(Z\alpha)^2}{\pi^2}\int\frac{f_1(k,m_1,m_2) d^4k}
{[(k-p)^2+i\epsilon][(k-q)^2+i\epsilon]D_e(k)D_\mu(-k)},
\end{equation}
\begin{displaymath}
f_1(k,m_1,m_2)=m_2(4m_1+2k_0)-2m_1k_0-2k^2_0+\frac{2}{3}\vec k^2,
\end{displaymath}
\begin{displaymath}
D_\mu(-k)=k^2-2E_2k_0+b^2+i\epsilon\approx -2m_2k_0+i\epsilon.
\end{displaymath}
\begin{equation}
V_{2\gamma}^{(b)}(\vec p,\vec q)=\frac{i(Z\alpha)^2}{\pi^2}\int \frac{f_2(k,m_1,m_2)d^4k}
{[(k-p)^2+i\epsilon][(k-q)^2+i\epsilon]D_e(k)D_\mu(p-q-k)},
\end{equation}
\begin{displaymath}
f_2(k,m_1,m_2)=m_2(4m_1+2k^0)-2m_1k_0-6k^2_0+\frac{10}{3}\vec p\vec k+\frac{10}{3}
\vec q\vec k+\frac{10}{3}\vec k^2,
\end{displaymath}
\begin{displaymath}
D_\mu(p-q-k)=k^2+2E_2k_0+2\vec k(\vec p+\vec q)-
(\vec p+\vec q)^2+b^2+i\epsilon\approx 2m_2k_0+i\epsilon.
\end{displaymath}

\begin{figure}
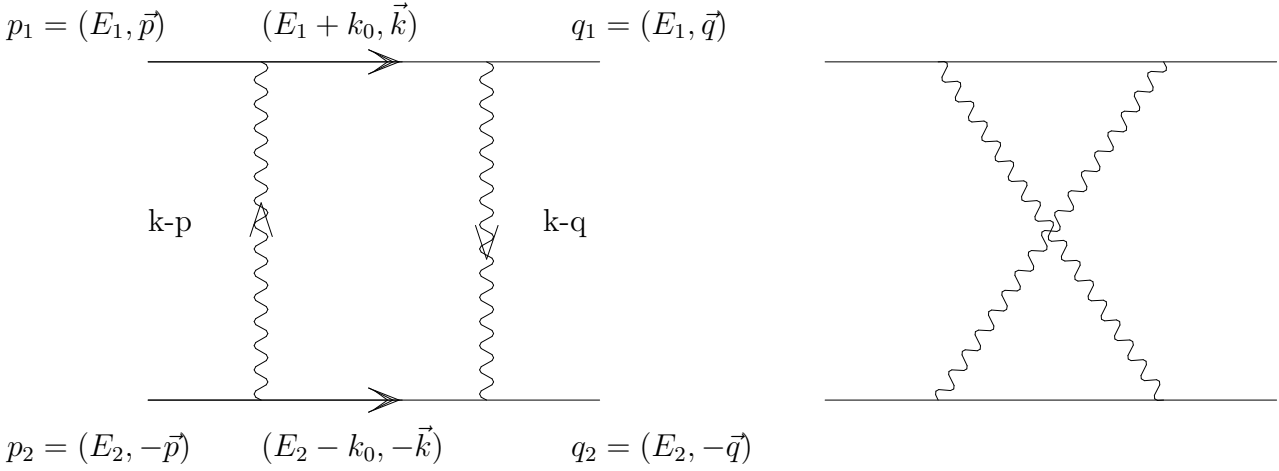

\magnitude=2000
\GRAPH(hsize=15){
\mov(1,0){\lin(4,0)}%
\mov(-0.25,-0.5){$p_2=(E_2,-\vec p)$}%
\mov(-0.25,3.25){$p_1=(E_1,\vec p)$}%
\mov(4.75,-0.5){$q_2=(E_2,-\vec q)$}%
\mov(4.75,3.25){$q_1=(E_1,\vec q)$}%
\mov(1,1.5){k-p}%
\mov(4.5,1.5){k-q}%
\mov(2,3.25){$(E_1+k_0,\vec k)$}%
\mov(2,-0.5){$(E_2-k_0,-\vec k)$}%
\mov(1,3){\lin(4,0)}%
\mov(7,0){\lin(4,0)}%
\mov(7,3){\lin(4,0)}%
\mov(1,0){\arrow(2.25,0)}%
\mov(1,3){\arrow(2.25,0)}%
\mov(2,0){\wavelin(0,3)}%
\mov(2,1.75){\lin(-0.1,-0.3)}%
\mov(2,1.75){\lin(0.1,-0.3)}%
\mov(4,1.25){\lin(-0.1,0.3)}%
\mov(4,1.25){\lin(0.1,0.3)}%
\mov(4,0){\wavelin(0,3)}%
\mov(8,0){\wavelin(2,3)}%
\mov(8,3){\wavelin(2,-3)}%
}

\vspace{10mm}

\caption{\bf Direct and crossed two-photon diagrams of exchange
interaction}
\end{figure}

\vspace{3mm}

The main contribution of $V_{2\gamma}$ in the energy spectrum is
proportional to $\alpha^5$. The analysis of corrections $O(Z\alpha)^6$
show, that they may exist in the energy levels also, if we take into account,
for example, the contribution of photon poles, when
$k_0\sim\alpha$, $|\vec p|\sim\alpha$, $|\vec q|\sim\alpha$, $|\vec k|\sim\alpha$.
In order to extract such terms from $V_{2\gamma}$ let transform the product of
electron and muon denominators in the direct two-photon diagram in the following
manner:
\begin{equation}
\frac{1}{D_e(k)D_\mu(-k)}=\frac{-2\pi i\delta(k_0)}{-2E(\vec k^2-b^2)}-
\frac{1}{2E}\left[\frac{1}{(k_0+i\epsilon)D_e(k)}+\frac{1}{(-k_0+i\epsilon)
D_\mu(-k)}\right],
\end{equation}
where the addendum with $\delta(k_0)$ in the right part of (30) cancels with
the iteration term of quasipotential. First addendum in the square brackets has
the same structure in the leading order over $1/m_2$ as a crossed amplitude.
Then the two-photon interaction quasipotential, which generates the correction
$(Z\alpha)^6m_1^2/m_2$ is equal:
\begin{equation}
V_{2\gamma}(\vec p,\vec q)=\frac{2i(Z\alpha)^2}{3\pi^2}\int\frac{d^4k[4\vec k^2+
5\vec k(\vec p+\vec q)-6k_0^2]}{[(k-p)^2+i\epsilon][(k-q)^2+i\epsilon]
D_e(k)(2m_2k_0+i\epsilon)}.
\end{equation}

We have calculated the contribution of $V_{2\gamma}$ to the energy spectrum for
n=1 and n=2 by means of system for analytical calculations "Mathematica" \cite{SW}
(package feynpar.m). The corresponding results are:
\begin{equation}
\Delta B_7=\left\{{-\frac{7m_1^2}{2m_2}(Z\alpha)^6,~~~~~n=1}
\atop{-\frac{31m_1^2}{16m_2}(Z\alpha)^6,~~~~~n=2}\right.
\end{equation}

\section{Three-photon exchange interaction}

There are six diagrams, shown on Fig.2, which determine the three-photon
exchange interaction in the muonium.

\begin{figure}
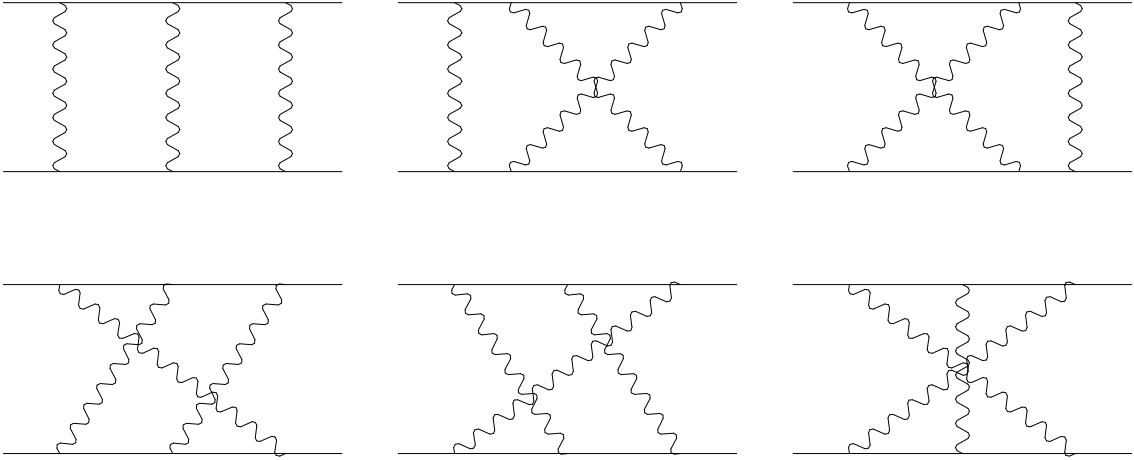

\magnitude=2000
\GRAPH(hsize=15){
\mov(0,2){\lin(3,0)}%
\mov(0,3.5){\lin(3,0)}%
\mov(3.5,2){\lin(3,0)}%
\mov(7,2){\lin(3,0)}%
\mov(3.5,3.5){\lin(3,0)}%
\mov(7,3.5){\lin(3,0)}%
\mov(0.5,2){\wavelin(0,1.5)}%
\mov(1.5,2){\wavelin(0,1.5)}%
\mov(4,-0.5){\wavelin(2,1.5)}%
\mov(4,1){\wavelin-(1,-1.5)}%
\mov(5,1){\wavelin-(1,-1.5)}%
\mov(2.5,2){\wavelin(0,1.5)}%
\mov(4,2){\wavelin(0,1.5)}%
\mov(4.5,2){\wavelin(1.5,1.5)}%
\mov(7.5,2){\wavelin(1.5,1.5)}%
\mov(7.5,3.5){\wavelin-(1.5,-1.5)}%
\mov(7.5,-0.5){\wavelin(2,1.5)}%
\mov(0.5,-0.5){\wavelin(1,1.5)}%
\mov(1.5,-0.5){\wavelin(1,1.5)}%
\mov(0.5,1){\wavelin-(2,-1.5)}%
\mov(7.5,1){\wavelin-(2,-1.5)}%
\mov(4.5,3.5){\wavelin-(1.5,-1.5)}%
\mov(8.5,-0.5){\wavelin(0,1.5)}%
\mov(9.5,2){\wavelin(0,1.5)}%
\mov(0,1){\lin(3,0)}%
\mov(0,-0.5){\lin(3,0)}%
\mov(3.5,1){\lin(3,0)}%
\mov(7,1){\lin(3,0)}%
\mov(3.5,-0.5){\lin(3,0)}%
\mov(7,-0.5){\lin(3,0)}%
}
\caption{\bf Feynman diagrams of three-photon exchange interaction in
muonium}
\end{figure}

Let consider the first diagram of Fig.2. The corresponding amplitude
already has the factor $\alpha^6$, which appears due to electromagnetic
vertices and Coulomb wave function. So, in the first stage of our
calculations we have neglected by the electron and muon vector momentum
of relative motion in the initial and final states, taking into account
that the necessary accuracy is already achieved. Then the first diagram
amplitude (Fig.2) takes the form:
\begin{equation}
T_1^{3\gamma}=-\frac{(Z\alpha)^3}{4\pi^5}\int d^4p\int d^4p'
\frac{<\gamma_1^\lambda(\hat q_1-\hat p'+m_1)\gamma_1^\nu(\hat p_1-\hat p+m_1)
\gamma_1^\mu>}{(p^2-w^2+i\epsilon)(p'^2-w^2+i\epsilon)[(p-p')^2+i\epsilon]}
\end{equation}
\begin{displaymath}
\frac{<\gamma_2^\mu(\hat p_2+\hat p+m_2)\gamma_2^\nu(\hat q_2+\hat p'+m_2)\gamma_2^\lambda>}
{D_e(p)D_e(p')D_\mu(-p)D_\mu(-p')},
\end{displaymath}
where $D_{e,\mu}(p)$ are the denominators of electron and muon propagators:
\begin{equation}
D(\pm p)=p^2-w^2\pm 2mp^0+i\epsilon,~~~w^2=-b^2,
\end{equation}
and the corner brackets designate the averaging over Dirac bispinors;
$p_1, p_2$ are the four-momenta of particles in the initial state;
$q_1, q_2$ are the particle four-momentum in the final state. As usually is,
the factor $Z\alpha$ emphasizes exchanging character of photon interaction
between particles. The exchange photon propagators were taken in Feynman
covariant gauge. As it famous, using of Coulomb gauge is the most natural
for exchange photons, because the Coulomb interaction dominates in the system
$(e^-\mu^+)$. Nevertheless, the equivalence of Coulomb and Feynman gauges
in the scattering approximation for the three-photon diagrams calculations
was shown in \cite{BYG}. To construct the quasipotential of the system
$(e^-\mu^+)$ with L=0 and J=1, that corresponds to $T^{3\gamma}_1$,
let introduce the projector operator for initial and final states of the
kind (6), setting also $\vec p=\vec q=0$.
Projecting the particles on $^3S_1$- state by means of (6), we avoid
cumbersome matrix multiplication in the bispinor averages and immediately
pass on to calculation of total trace in (33). As a result, the quasipotential
of first diagram may be written in the form:
\begin{equation}
V_1^{3\gamma}=-\frac{(Z\alpha)^3}{\pi^5}\int d^4p\int d^4p'\frac{F_1(p,p')}
{D_\gamma(p)D_\gamma(p')D_\gamma(p-p')D_e(p)D_e(p')D_\mu(-p)D_\mu(-p')},
\end{equation}
$D_\gamma(p)=p^2-w^2+i\epsilon$,
\begin{equation}
F_1(p,p')=f_{12}(p,p')m_2^2+\frac{1}{3}f_{11}m_2,~~
f_{12}=pp'-4m_1^2-2m_1p_0-2m_1p_0'-2p_0p_0',
\end{equation}
\begin{displaymath}
f_{11}(p,p')=2m_1p'^2+p_0p'^2+10m_1pp'+2p_0pp'+2p_0'pp'+
\end{displaymath}
\begin{displaymath}
2m_1p^2+p_0'p^2+6m_1^2p_0+6m_1^2p_0'+4m_1p_0^2+4m_1p_0'^2-4m_1p_0p_0'.
\end{displaymath}

We kept in (35) only the terms proportional to $m^2_2$ and $m_2$, taking in mind
the determination of contribution to muonium fine structure in the leading
order over parameter $m_1/m_2$. As will soon become evident, we can't restrict
in $F(p,p')$ only by terms $\sim m_2^2$. The quasipotentials of the rest
amplitudes of Fig.2 may be constructed in a similar way. They differ from
each other due to momentum dependence in muonic denominators and to the kind
of functions $f_{i1}$ (i=1,...,6).

The parts of $F_i(p,p')$, proportional to $m_2^2$, coincide in all six
amplitudes. Let remark, that when substitute $\hat\epsilon\rightarrow\gamma_5$
in projector operator (6) ($^1S_0$ state), we obtain the same function
$f_{12}(p,p')$, as for the $^3S_1$ muonium. This means, that the muonium
hyperfine splitting appears as effect of higher order on $m_1/m_2$.
Functions $f_{i1}$ are equal to:
\begin{equation}
f_{21}=-10m_1p'^2-5p_0p'^2+10m_1pp'+4p_0pp'-4p'_0pp'+2m_1p^2+4p'_0p^2+
12p_0m_1^2-
\end{equation}
\begin{displaymath}
-6m_1^2p'_0+4m_1p_0^2+8m_1p_0p'_0-8m_1p'^2_0+4p'_0p_0^2,
\end{displaymath}
\begin{equation}
f_{31}=2m_1^2p'^2+4p_0p'^2+10m_1pp'-4p_0pp'+4p'_0pp'-10m_1p^2-5p'_0p^2-
\end{equation}
\begin{displaymath}
-6m_1^2p_0+12m_1^2p'_0-8m_1p_0^2+8m_1p_0p'_0+4m_1p'^2_0+4p_0p'^2_0,
\end{displaymath}
\begin{equation}
f_{41}=2m_1p'^2+p_0p'^2-2m_1pp'+4p_0pp'+2p'_0pp'-10m_1p^2-8p'_0p^2-12m_1^2p_0+
\end{equation}
\begin{displaymath}
+6m_1^2p'_0+4m_1p_0^2-4m_1p_0p'_0+4m_1p'^2_0-8p'_0p^2_0,
\end{displaymath}
\begin{equation}
f_{51}=-10m_1p'^2-8p_0p'^2-2m_1pp'+2p_0pp'+4p_0'pp'+2m_1p^2+p'_0p^2+6m_1^2p_0-
\end{equation}
\begin{displaymath}
-12m_1^2p'_0+4m_1p^2_0-4m_1p_0p'_0+4m_1p'^2_0-8p'^2_0p_0,
\end{displaymath}
\begin{equation}
f_{61}=-10m_1p'^2-5p_0p'^2-2m_1pp'-4p_0pp'-4p'_0pp'-10m_1p^2-5p'_0p^2-
\end{equation}
\begin{displaymath}
-6m_1^2p_0-6m_1^2p'_0-8m_1p^2_0-4m_1p_0p'_0-8m_1p'^2_0.
\end{displaymath}

Integral function in (35) has simple poles over loop energies $p_0, p'_0$ in
electron, muon and photon propagators. So, the most natural way of
integration (35) consists in the calculation of integrals on $p_0, p'_0$
at the first step, using the method of residues. But such an approach of
calculation leads, nevertheless, to rather complicated intermediate
expressions, what makes highly questionable its subsequent analytical
integration on spatial momenta $\vec p, \vec p'$. So, we have used different
approach of integration in (4), connected with transformation of muonic
denominators, accounting the necessary calculational accuracy on parameter
$m_1/m_2$. Considering that the spatial momentum of muonic motion in the
intermediate state $|\vec p|< m_2$, we obtain:
\begin{equation}
D_\mu(p)=p^2-w^2+2m_2 p_0\approx 2m_2\left(p_0-\frac{\vec p^2+w^2}{2m_2}+
i\epsilon\right)\approx 2m_2(p_0+i\epsilon),
\end{equation}
where the second approximate equality means that we neglect by the muon
kinetic energy in the intermediate state. Doing so, we suppose that the
integration contour on variable $p_0$ must be closed in the lower halfplane.
Considering the terms, proportional to $m_2^2$ in the numerators of all six
diagrams (function $f_{12}(p,p')$), we have arrived to the need of expression
transformation, which includes the sum of muonic denominators (34). Using the
second approximate equality from (42), we obtain:
\begin{displaymath}
\frac{1}{D_\mu(-p)D_\mu(-p')}+\frac{1}{D_\mu(-p)D_\mu(p'-p)}+\frac{1}{D_\mu(-p')D_\mu(p-p')}+
\end{displaymath}
\begin{displaymath}
+\frac{1}{D_\mu(p)D_\mu(p-p')}+\frac{1}{D_\mu(p')D_\mu(p'-p)}+\frac{1}{D_\mu(p')D_\mu(p)}\approx
\end{displaymath}
\begin{equation}
\approx\frac{(-2\pi i)\delta(p_0)}{2m_2}\frac{(-2\pi i)\delta(p'_0)}
{2m_2}.
\end{equation}

In the energy spectrum the expression (43) will cause the corrections of
order $O(\alpha^4)$, which are canceled by the similar terms from the
iteration part of the quasipotential. Consequently, to find the necessary
contribution of order of $\alpha^6$, we must use first approximate equality
in (42). Taking the difference
\begin{equation}
\frac{1}{2m_2\left(p_0-\frac{\vec p^2+w^2}{2m_2}+i\epsilon\right)}-
\frac{1}{2m_2(p_0+i\epsilon)}\approx\frac{(\vec p^2+w^2)}{4m^2_2(p_0+i\epsilon)^2},
\end{equation}
let represent the quantity $1/D_\mu(p)$ in the form:
\begin{equation}
\frac{1}{D_\mu(p)}\approx\frac{1}{2m_2(p_0+i\epsilon)}+\frac{(\vec p^2+w^2)}
{4m^2_2(p_0+i\epsilon)^2}.
\end{equation}
Second addendum of (45) is of higher order on $m_1/m_2$ in comparison with the
first. But it leads to the necessary order correction  on the other
parameter $\alpha$. Using the splitting (45), in the sum (43), let extract the terms,
which generate the correction $O(\alpha^6)$ and $O(\alpha^6\ln\alpha)$ in
the energy spectrum. We may write them in the following manner:
\begin{equation}
\frac{(\vec p'^2+w^2)}{8m^3_2}\left[\frac{2\pi i\delta(p'_0)}{(p_0+i\epsilon)^2}-
\frac{2\pi i\delta(p'_0-p_0)}{(p_0+i\epsilon)^2}-\frac{2\pi i\delta(p_0)}{(p'_0+
i\epsilon)^2}\right]+
\end{equation}
\begin{displaymath}
+\frac{(\vec p^2+w^2)}{8m^3_2}\left[\frac{-2\pi i\delta(p'_0)}{(p_0+i\epsilon)^2}-
\frac{2\pi i\delta(p'_0-p_0)}{(p_0+i\epsilon)^2}+\frac{2\pi i\delta(p_0)}{(p'_0+
i\epsilon)^2}\right]+
\end{displaymath}
\begin{displaymath}
+\frac{(\vec p-\vec p')^2+w^2}{8m^3_2}\left[-\frac{2\pi i\delta(p'_0)}{(p_0+i\epsilon)^2}+
\frac{2\pi i\delta(p'_0-p_0)}{(p_0+i\epsilon)^2}-\frac{2\pi i\delta(p_0)}{(p'_0+
i\epsilon)^2}\right]
\end{displaymath}

It is evident from three-photon interaction amplitude of the type (33), that
the parts of (46) give the necessary order corrections on $\alpha$ in the studied
fine structure intervals. The same order corrections $O(m_1/m_2)$, as well as (18),
will arise from the quasipotential terms containing the functions
$f_{i1}(p,p')$, when we use the second approximation of (42) for muonic
denominators. To do definite conclusion about the order of appearing terms
in energy spectrum, which are determined by these quasipotential addenda, let
transform them for greater simplification. Let consider for definiteness massless
terms in the function $f_{i1}(p,p')$, proportional to $\sim p^2, p'^2, pp'$:
\begin{equation}
3p^2\left[\frac{1}{D_\mu(p)}+\frac{1}{D_\mu(-p)}-\frac{1}{D_\mu(p-p')}-
\frac{1}{D_\mu(p'-p)}\right]+
\end{equation}
\begin{displaymath}
+3p'^2\left[\frac{1}{D_\mu(-p')}+\frac{1}{D_\mu(p')}-\frac{1}{D_\mu(p'-p)}-
\frac{1}{D_\mu(p-p')}\right]-
\end{displaymath}
\begin{displaymath}
-6pp'\left[\frac{1}{D_\mu(-p')}+\frac{1}{D_\mu(p')}+\frac{1}{D_\mu(-p)}+
\frac{1}{D_\mu(p)}-\frac{1}{D_\mu(p'-p)}-\frac{1}{D_\mu(p-p')}\right]\approx
\end{displaymath}
\begin{displaymath}
\approx\frac{3p^2}{2m_2}\left[-2\pi i\delta(p_0)+2\pi i\delta(p_0-p'_0)\right]+
\frac{3p'^2}{2m_2}\left[-2\pi i\delta(p'_0)+2\pi i\delta(p_0-p'_0)\right]-
\end{displaymath}
\begin{displaymath}
-\frac{6pp'}{2m_2}\left[-2\pi i\delta(p_0)-2\pi i\delta(p'_0)+2\pi i\delta(p_0-p'_0)\right].
\end{displaymath}
The transformation of other terms in functions $f_{i1}(p,p')$ may be carried
out by analogy. The next period of calculation consists in the integration over
four-momenta in expressions (46)-(47). The typical two-loop integral, that
results on this way has the following structure \cite{BYG}:
\begin{equation}
K_i=(4\pi)^2\int\frac{d^4p d^4p'}{-(2\pi)^8}\frac{G_i(p'_0,p_0,m_1)P(\vec p,
\vec p',w)}{(p'^2-w^2+i\epsilon)[(p-p')^2+i\epsilon](p^2-w^2+i\epsilon)D_e(p')
D_e(p)},
\end{equation}
where $G_i(p'_0,p_0,m_1)$ contains one $\delta$ - function, and
$P(\vec p', \vec p, w)$ is a polynomial. To calculate fundamental integrals (48)
we have used  Feynman parameterization in order to combine the denominators
of the particle propagators, and the symmetry properties of the integral with
the replacement $p\Leftrightarrow p'$. There is the next set of functions
$G_i(p'_0,p_0,m_1)$, which appear in this paper:
\begin{equation}
G_1=-\frac{2\pi i\delta(p_0-p'_0)2m_1}{(p_0+i\epsilon)^2},~~~
G_2=-\frac{2\pi i\delta(p_0) 2 m_1}{(p'_0+i\epsilon)^2},~~~
G_3=-2\pi i\delta(p_0) 2m_1,
\end{equation}
\begin{displaymath}
G_4=-2\pi i\delta(p'_0)2m_1,~~~G_5=-2\pi i\delta(p_0-p'_0)2m_1.
\end{displaymath}
The results of the integrations for $K_i$ (48) are presented in the table
\cite{BYG}.\\[8mm]
{\bf Table of the integrations $K_i$ (48), appearing in the\\ muonium fine
structure calculations}\\[3mm]
\begin{tabular}{|c|c|c|c|c|} \hline
    & $\vec p^2(\vec p\vec p')$ & $(\vec p\vec p')^2$  & $\vec p'^2(\vec p\vec p')$ & $ w^2(\vec p\vec p') $	\\  \hline
$K_1$ & $2\ln 2-\frac{1}{2}$ & $2\ln 2-\frac{1}{2}$ & $2\ln 2-\frac{1}{2}$ & 0			       \\     \hline
      &$\vec p^2(\vec p'^2-\vec p\vec p')$ & $\vec p\vec p'(\vec p'^2-\vec p\vec p')$ & $w^2(\vec p'^2-\vec p\vec p')$ & $w^2\vec p^2$	     \\      \hline
$K_2$ &$\frac{1}{2}\ln\frac{m_1}{2w}-\frac{1}{32}$ &$\frac{1}{4}\ln\frac{m_1}{2w}-\frac{13}{32}$ &$\frac{5}{32}$   & $\frac{2}{3}$	      \\      \hline
      &  $\vec p'^2$  & $\vec p\vec p'$   & $\vec p^2 $ &    $w^2$  \\	\hline
$K_3$ &    ---	    & $\frac{1}{4}\ln\frac{m_1}{2w}-\frac{1}{4}$  &  $ \frac{1}{2}\ln\frac{m_1}{2w}-\frac{1}{8}$   &   $\frac{1}{8}$  \\  \hline
$K_4$ & $\frac{1}{2}\ln\frac{m_1}{2w}-\frac{1}{8}$ &  $\frac{1}{4}\ln\frac{m_1}{2w}-\frac{1}{4}$ &---	&  $\frac{1}{8}$   \\  \hline
 $K_5$& $\ln 2$ & $\ln 2-\frac{1}{2}$ & $\ln 2$ & 0 \\	  \hline
\end{tabular}
\vspace{5mm}

Then the contributions, defined by expressions (37-41) and (46) are correspondingly
equal:
\begin{equation}
\Delta B_1=-\frac{1}{2}(Z\alpha)^6\frac{m_1^2}{m_2}
\end{equation}
\begin{equation}
\Delta B_2=(Z\alpha)^6\frac{m^2_1}{m_2}(6\ln 2-\frac{11}{48})
\end{equation}

We have introduced in (48) the photon mass w to avoid "infrared" singularities.
The "infrared" logarithms $\ln w$, containing this photon mass (see table of
integrals $K_i$), and appearing at intermediate expressions, are mutually
cancelled in the corrections $\Delta B_1$, $\Delta B_2$.

Let consider now the quasipotential addenda, containing the momenta of
particle relative motion in the initial and final states. We denote them by
$\vec r_1$ and $\vec r_2$ correspondingly.
Their consideration leads to modification of $f_{i1}$, which acquire the
following additional terms:
\begin{equation}
\Delta f_{21}=10m_1p'r_2+5p_0p'r_2+m_1pr_2+3p'_0pr_2,
\end{equation}
\begin{equation}
\Delta f_{31}=-m_1p'r_1-3p_0p'r_1-10m_1pr_1-5p'_0pr_1,
\end{equation}
\begin{equation}
\Delta f_{41}=m_1(7p'r_1+5p'r_2+10pr_1+5pr_2)+
6p_0p'r_1+5p_0p'r_2+8p'_0pr_1+5p'_0pr_2,
\end{equation}
\begin{equation}
\Delta f_{51}=m_1(-5p'r_1-10p'r_2-5pr_1-7pr_2)-5p_0p'r_1-8p_0p'r_2-5p'_0pr_1-6p'_0pr_2,
\end{equation}
\begin{equation}
\Delta f_{61}=m_1(-11p'r_1-11p'r_2-11pr_1-11pr_2)-
8p_0p'r_1-8p_0p'r_2-8p'_0pr_1-8p'_0pr_2.
\end{equation}
Using again the symmetry properties of appearing integrals under simultaneous
variable replacement $p\Leftrightarrow p'$, $r_1\Leftrightarrow r_2$, we
obtain cancellation of all integrations in (52)-(56). So, the contribution
of the particle relative motion in the fine structure with the accuracy
$O(m_1/m_2)$ is equal to zero. Thus the full value of the calculated
correction $(Z\alpha)^6 m_1^2/m_2$ for hydrogen-like system S-states is
defined as a sum of expressions (50) and (51):
\begin{equation}
\Delta B_8=(Z\alpha)^6\frac{1}{n^3}\frac{\mu^3}{m_1m_2}\left(6\ln 2-\frac{35}{48}\right).
\end{equation}

\section{Discussion of the results}

In this paper we have calculated all possible corrections of order
$(Z\alpha)^6m_1^2/m_2$ in the fine structure of hydrogen-like system on
the basis of diagrammatic approach. Our total result is determined by the
sum of terms $\Delta B_i$ (4), (8), (11), (13), (24), (27), (32) and (57):
\begin{equation}
\Delta B_{tot.}=\left(\frac{91}{12}\ln 2-\frac{545}{144}-\frac{17}{6n}-
\frac{37}{36n^2}+\frac{187}{96n^3}\right)
\frac{m_1^2(Z\alpha)^6}{m_2n^3}+\varepsilon_n
\end{equation}
\begin{displaymath}
\varepsilon_n=\left\{{-\frac{23m_1^2(Z\alpha)^6}{12m_2},~~~~~n=1,
\atop -\frac{31m_1^2(Z\alpha)^6}{128m_2},~~~~~n=2}\right.
\end{displaymath}

Numerical value of obtained contribution (58) for the "large" muonium fine
structure interval $2^3S_1\div 1^3S_1$ is equal to 0,19 MHz. Recently, the
calculation of the recoil corrections $O(Z\alpha)^6m_1^2/m_2$ for the S-levels
of hydrogen atom was done in \cite{PG,Y,EG}. The total contribution of the
necessary order in the energy spectrum, which was obtained by using Braun
formula, is equal \cite{EG}:
\begin{equation}
\Delta E_{tot.}=\left(\frac{1}{8}+\frac{3}{8n}-\frac{1}{n^2}+\frac{1}{2n^3}\right)
\frac{(Z\alpha)^6m_1^2}{m_2}+\left(4\ln 2-\frac{7}{2}\right)\frac{(Z\alpha)^6m_1^2}
{m_2}.
\end{equation}
Comparing results (58) and (59), we see that the analytical expression of our
correction (58) differs from (59), because we have calculated the contribution
of necessary order to the energy spectrum, coming both from spin-independent
and spin-dependant parts of the quasipotential.
Numerical values of the contributions (58) and
(59) for the levels with n=1 and n=2 are equal correspondingly:
n=1: -0,212 MHz, -0,065 MHz; n=2: -0,021 MHz, -0,006 MHz. The contribution
of correction (58) to the fine structure interval $2S\div 1S$ of hydrogen atom
is equal to 21,5 KHz, whereas the values of similar contributions, obtained
in \cite{EG,E} are equal correspondingly: 6,6 KHz \cite{EG}; 14,5 KHz \cite{E}.

We are grateful to I.B. Khriplovich, S.G. Karshenboim, E.A. Kuraev, P.Labelle,
V.A.. Saleev, R.A. Sen'kov, A.S. Yelkhovsky for useful discussions. We thank
H. Grotch for the copy of paper \cite{PG} and R. Mertig for the description
of FeynCalc package. The work was performed under the financial support of the Russian
Foundation for Fundamental Research (grant 98-02-16185) and the Program
"Universities of Russia - Fundamental Researches" (grant 2759).

\end{document}